# Theoretical study on copper's energetics and magnetism in TiO$_2$ polymorphs

M. Hussein. N. Assadi[*]

*School of Materials Science and Engineering, University of New South Wales, Sydney, NSW 2052, Australia*

Dorian A. H. Hanaor

*Particles and Grains Laboratory, School of Civil Engineering, University of Sydney, NSW 2006, Australia*

[*]Corresponding Author: email hussein.assadi@unsw.edu.au; tel: +61-2-93855234

**Abstract:**

Density functional theory calculations were employed to model the electronic structure and the magnetic interactions in copper doped anatase and rutile titanium dioxide in order to shed light on the potential of these systems as magnetic oxides using different density functional schemes. In both polymorphs copper dopant was found to be most stable in substitutional lattice positions. Ferromagnetism is predicted to be stable well above room temperature with long range interactions prevailing in the anatase phase while the rutile phase exhibits only short range superexchange interaction among nearest-neighbour Cu ions. Additionally, energetic evaluation of dopants in scattered and compact configurations reveals a dopant clustering tendency in anatase TiO$_2$.

Keywords: DFT, TiO$_2$, DMS, Ab-initio, Copper doped, Spintronics

PACS: 31.15.A-, 77.84.Bw, 61.72.uj

# I. INTRODUCTION

Magnetically doped oxide systems have attracted considerable attention because of their potential ferromagnetic behaviour at room temperature[1] which can be utilized for various spintronic devices facilitating the injection, switching and amplification of spin-polarized current.[2,3] In these systems, diluted concentrations of transition metal (TM) elements are doped in the host lattice with the aim of achieving intrinsic room temperature ferromagnetism. Ferromagnetism among TM ions is mediated by two mechanisms (a) itinerant magnetism that prevails in the presence of mobile carriers,[4] and (b) by bound magnetic polarons that prevail in ionic systems.[5]

Doped $TiO_2$ systems have been previously studied as diluted magnetic oxides (DMOs).[6] $TiO_2$ is of particular interest owing to its multifunctional properties and its existence in two easily obtained kinetically stable polymorphs.[7] However, insufficient understanding of the TM dopants' electrical and structural behaviour within the host oxide impedes the ability to engineer room temperature $TiO_2$-based DMOs. Sensitivity of the system's magnetic properties to preparation methods and consequent levels of intrinsic and extrinsic defects[8] has caused an ongoing controversy with respect to the origin of magnetism and the mechanisms involved.[9] In particular, uncertainty exists regarding the segregation of magnetic dopants that may lead to the formation of dispersed secondary ferromagnetic phases.[10] For instance, in Cobalt doped anatase $TiO_2$ prepared by sputtering in pure Ar, a uniform Co distribution was reported,[11] while for samples prepared by oxygen plasma assisted molecular beam epitaxy, Co rich droplets were detected on the surface[12] (Full review at Ref. 13). Dispersed nano-scale secondary magnetic phases are often not readily detected by conventional micro-analytical methods (XRD, TEM), and this may result in false positive identification of intrinsic magnetism.

To circumvent experimental difficulties, density functional theory (DFT) calculations are frequently employed to assess DMOs.[14,15] However, after a decade, this approach has yet to produce conclusive analyses of magnetism in TM oxides.[16] The main problem of conventional DFT calculations based on generalized gradient approximation (GGA) and local density approximation (LDA) is the underestimation of the band gap. Earlier LDA[17] and GGA[18] calculations have predicted bandgap values of ~2 eV for rutile $TiO_2$ which is ~30% smaller than experimental values.[19] Bad-gap underestimation in DMOs leads to false predictions of carrier mediated magnetism, as impurity levels are predicted to lie in the conduction band. The shortcomings of LDA and GGA formalism based DFT can be greatly improved by the appropriate use of a Hubbard term, U, i.e. GGA+U formalism.[20] By accounting for the repulsion and exchange of 3$d$ electrons, a careful choice of U allows an improved description of the band gap, thus allowing the correct prediction of magnetic behaviour.[21] Previous GGA+U studies have focused on the rutile phase of $TiO_2$ doped with TM elements[22] and the role of its native defects[23] and offer no comparative description of the metastable anatase phase that is often realized in the laboratory conditions.

Here we report the study of the electronic and magnetic behaviour of Cu doped in both anatase and rutile $TiO_2$ using DFT. The significance of $TiO_2$:Cu as a DMO system lies in the fact that neither metallic copper nor any of its oxides exhibit ferromagnetism. Consequently, any observation of ferromagnetic behaviour in the system can only be interpreted to be intrinsic. Here the formation energies and local charge populations of Cu dopants in both anatase and rutile $TiO_2$ are investigated using two DFT formalisms; (a) all electronic calculations based on non-local exchange-correlation functional within the framework of general gradient approximations (GGA) using Dmol3 code and (b) planewave-pseudopotential calculations based on orbital-dependant GGA+U formalism using VASP. Furthermore, the present work investigates the effects of dopants' different charge states on magnetic interactions.

## II. METHODS

To probe the energetics of Cu dopants in $TiO_2$ host lattices, we employed two different DFT methods. First, the formation energies of both substitutional and interstitial Cu in rutile (r-$TiO_2$) and anatase (a-$TiO_2$) were calculated using DMol3 code with generalized gradient approximation (GGA) for the exchange-correlation functional.[24, 25] Full details and justifications of the computational settings have been reported previously.[26] The advantage of the DMol3 method is its all-electronic approach in which both valence and core electrons are explicitly considered in the calculation of the total energy. This is as opposed to the use of pseudo-potentials for the core electrons as in many DFT formalisms. However, advanced non-local functionals for exchange-correlation energy have not yet been implemented in DMol3. This might result in inaccuracies in placing the d-shell electrons with respect to the band edges.

To further examine the obtained DMol3 results and also provide a comparative study, We also calculated the formation energies of copper dopants using Vienna ab initio simulation package (VASP),[27] using projector-augmented waves for expressing the electronic densities and Perdew–Burke–Ernzerhof functional for approximating the exchange-correlation functional.[28] VASP simulations were conducted using high precision settings for electronic minimization and the thresholds for geometry optimization were the same as with DMol3 calculations, $10^{-5}$ eV for energy and $10^{-3}$ eV/Å for the Cartesian components of the forces acting on the ions.

To circumvent the shortcomings of GGA, i.e. underestimation of the electronic band-gap, we applied an on-site, orbital dependent Hubbard term ($U$) to correct the position of the d-shell electrons with respect to band gap edges. A local Hubbard term of $U = 4.2$ eV was introduced to act on d-shell electrons of Ti and Cu as previously reported for these elements.[29, 30] It should be noted that introducing a phenomenological parameter such as $U$, renders the approach non-first-principles.

In both approaches, the primitive cells of pristine anatase and rutile were fully relaxed to their theoretical lattice parameters. Supercells with dimensions $3a \times 2b \times 2c$ relative to the primitive rutile and anatase unit cells were then constructed using these lattice values. The supercells for both anatase and rutile, contained 72 ions as illustrated in Fig. 1. For substitutional configurations one dopant atom per supercell was introduced. This results in a cation dopant concentration of 4.16 at%. After doping, internal coordinates of all ions in the supercell were allowed to relax, while maintaining calculated lattice parameters.

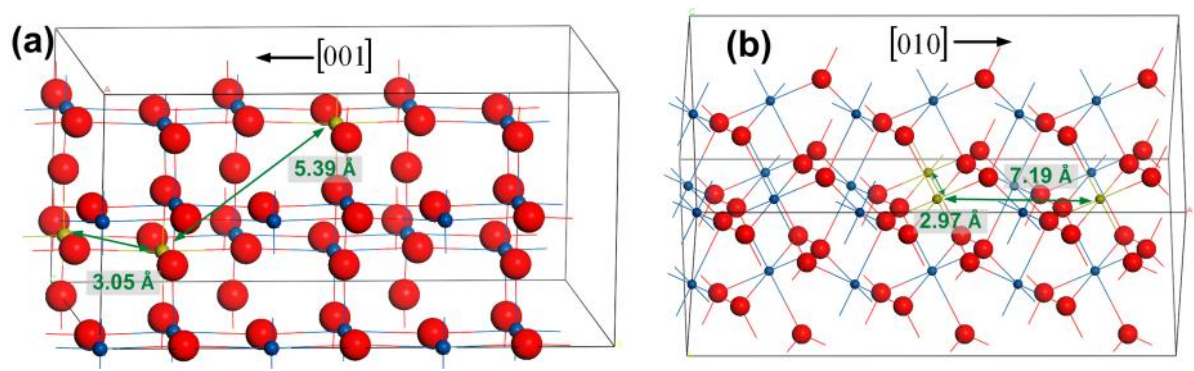

Fig. 1 (Color Online). (a) Anatase supercell and (b) rutile supercell showing copper dopant in scattered and compact configurations. The blue, gold and red spheres represent Ti, Cu and O respectively.

In the case of charged dopants in this work, for the sake of consistency, the charge states that are assigned to the dopants refer to the system's total charge. These values may differ from values expressed in conventional Kröger–Vink notation which usually applied to describe doped ionic systems. For instance $Cu_{Ti}^{2-}$ indicates a substitutional copper dopant, which has accepted two holes incurring a system charge of -2, while $Cu_{Int}^{2+}$ indicates an interstitial copper ion that has lost two electrons resulting in an overall system charge of +2. Assigning system charges allows the study of multiple charge states of electrically active dopants while maintaining supercell stoichiometry. To preserve charge neutrality as mandated by DFT, a uniform charge jellium is added to balance dopant charges.

Finally, magnetic interactions between copper ions in the host lattice were investigated by varying dopant position in the supercell. For this purpose two copper ions, creating 8.33% of Cu concentration, were substituted at Ti lattice positions at two different distances, creating scattered and compact dopant configurations. In the compact configurations copper ions are separated by a single oxygen ion and thus have the minimal possible distance from one another, which is 3.05 Å in anatase and 2.97 Å in rutile $TiO_2$. In the scattered configurations the copper ions take the maximal possible separation permitted by the supercell size, that is 5.38 Å in anatase and 7.19 Å in rutile $TiO_2$. Subsequently, the total energy was calculated once when the spins of the unpaired d-shell electrons of the copper ion dopant were set as parallel ($E_{FM}$) and once when these were set as anti-parallel ($E_{AFM}$). The difference of these two values ($\Delta E = E_{AFM} - E_{FM}$) is an indication of magnetic phase stability in the system, with larger $\Delta E$ values indicating a more stable ferromagnetic phase.

### III. RESULTS AND DISCUSSION

#### A. The formation energy of Cu in $TiO_2$ using GGA functional

In a-$TiO_2$, as shown in Fig. 2(a), for values of $E_{Fermi}$ near the valence band maximum (VBM), the interstitial $Cu_{Int}^{2+}$ is more stable than the substitutional $Cu_{Ti}^{-}$, having $E^f$ of 0.97 eV while the latter has an $E^f$ of 1.55 eV. As the Fermi level moves towards higher values, the stability of $Cu_{Ti}^{-}$ exceeds that of $Cu_{Int}^{2+}$. Both forms of interstitial and substitutional Cu experience multiple charge transitions within the permissible $E_{Fermi}$ levels. $Cu_{Ti}^{-}$ transitions to $Cu_{Ti}^{2-}$ at $E_{Fermi} = 0.14$ eV and then to $Cu_{Ti}^{3-}$ at $E_{Fermi} = 0.94$ eV. On the other hand $Cu_{Int}^{2+}$ transitions to $Cu_{Int}^{+}$ at $E_{Fermi} = 0.78$ eV and then to $Cu_{Int}^{\times}$ at $E_{Fermi} = 1.55$ eV. As interstitial doping would introduce donors and preclude $E_{Fermi}$ levels close to the VBM, the greater stability of $Cu_{Int}$ near VBM is only hypothetical. When we additionally consider the natural tendency of $TiO_2$ to possess oxygen vacancies and consequent intrinsic n-type behavior, with $E_{Fermi}$ levels close to conduction band minimum (CBM) at ~ 3eV, the stability of substitutional Cu dopant predominates, in the 3- charge state.

The electronic behavior of Cu in r-TiO$_2$ as in Fig. 2(b) resembles that in a-TiO$_2$. For $E_{Fermi}$ values near VBM, $Cu_{Int}^{2+}$ has a $E^f$ of -0.66 eV while the substitutional Cu has a charge state of 1+ with $E^f$ of 0.90 eV. However $Cu_{Ti}$ loses it high charge state by undergoing rather close transitions. For instance, at $E_{Fermi}$ = 0.27 eV, 0.40 eV and 0.67 eV it transitions to charge states of 0, 1-, and 2- respectively. Finally at $E_{Fermi}$ = 1.32 eV $Cu_{Ti}$ stabilizes at 3- which it holds for the rest of the band gap. As in the case of anatase, it can be assumed the substitutional Cu, in the 3- charge state, is the most stable configuration for experimentally attainable Fermi levels e.g. $E_{Fermi}$ =~ 3 eV.

At CBM level, $Cu_{Ti}^{3-}$ in a-TiO$_2$ has a $E^f$ of -6.97 eV while in r-TiO$_2$, it has a $E^f$ of -6.03 eV. This implies that Cu is slightly more soluble in the anatase phase than in rutile.

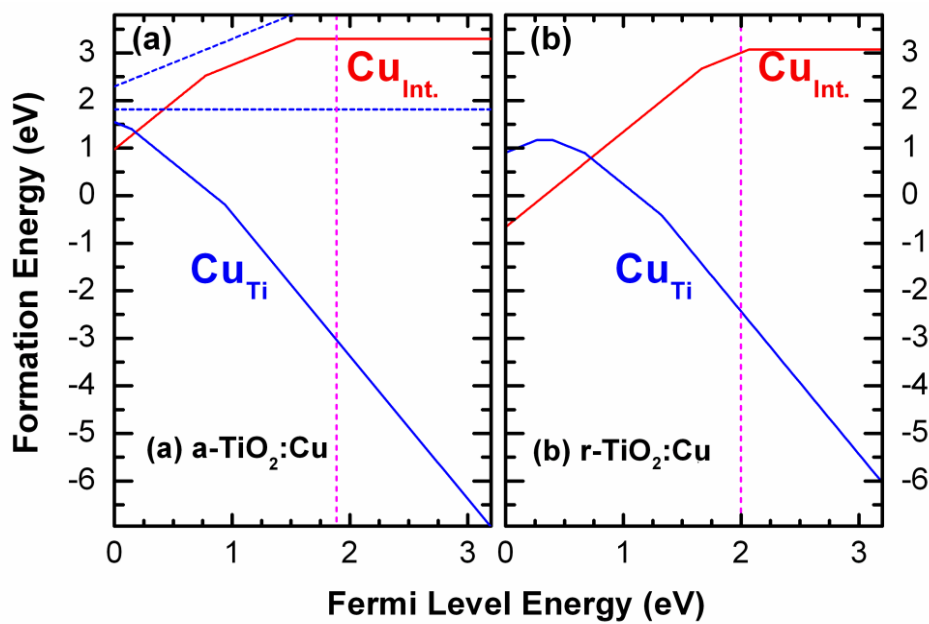

Fig. 2 (Color Online). The formation energy as a function of Fermi level for substitutional and interstitial Cu in both anatase (a) and rutile (b) TiO2 presented in the range of the experimental bandgap of 3.2 eV. The origin of the Fermi level axis is assumed to be at the valence band maximum. The dotted vertical pink lines indicate the position of the theoretical conduction band minimum. The Dotted blue lines represent those charge states that are never stable with respect to alternative charge states of substitutional Cu.

### B. GGA+U based evaluation of formation energy of Cu in TiO2

VASP prediction of formation energy for substitutional copper dopants in anatase and rutile TiO$_2$ is shown in Fig. 3. Results obtained using DMol3 code demonstrated that interstitial Cu has relatively higher formation energy than substitutional, therefore we did not consider $Cu_{Int}$ for analysis in this section. In a-TiO$_2$, as shown in Fig. 2(a), when $E_{Fermi}$ is located at the VBM, $Cu_{Ti}^+$ has a $E^f$ of -0.60 eV. As the Fermi level moves towards higher values, Cu experiences a double charge transition of ε(1+/1-) at $E_{Fermi}$ = 2.42 eV. As a result $Cu_{Ti}$ is never stable at a neutral charge state and forms a negative U centre in a-TiO$_2$. Consequently, Cu dopants act as donors for $E_{Fermi}$ values near VBM and as acceptors near CBM. This behaviour constitutes a charge-compensating phenomenon where the Cu dopants tend to neutralize other electrically active defects.

In doped rutile TiO$_2$, as shown in Fig. 3(b), Cu always stabilizes at 1+ charge state regardless of $E_{Fermi}$. For instance, for $E_{Fermi}$ = CBM, the $E^f$ of $Cu_{Ti}^+$ is 0.73 eV while $E^f$ of $Cu_{Ti}^\times$ is 1.45 eV. As a result Cu in rutile TiO$_2$ always acts as an electron donor. Furthermore, for $E_{Fermi}$ values near CBM as the case of n-type TiO$_2$, the formation energy of Cu dopant is 0.73 eV and 1.05 eV for anatase and rutile TiO$_2$ respectively, implying that Cu is more soluble in anatase TiO$_2$.

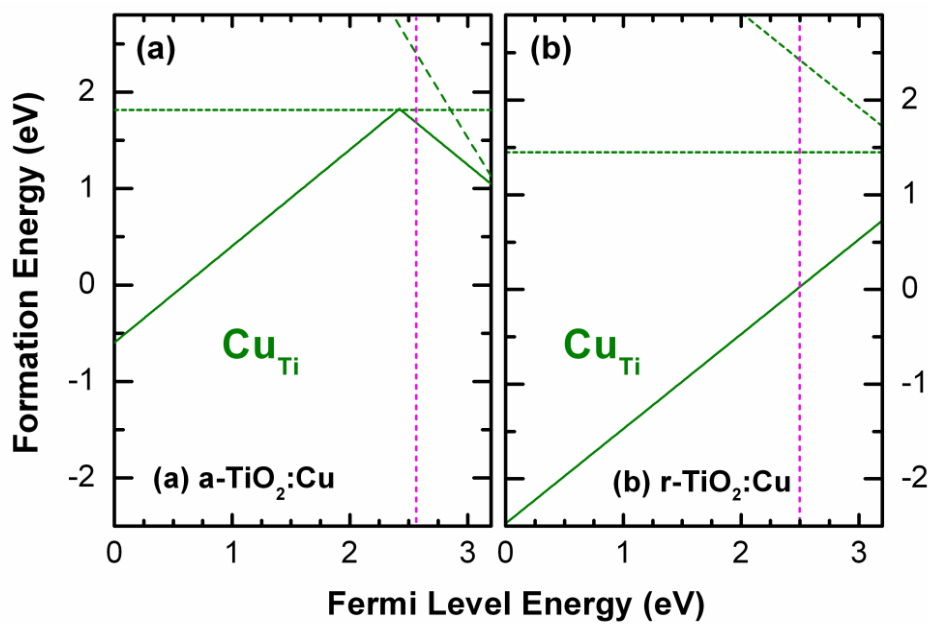

Fig. 3 (Color Online). Formation energy in substitutionally Cu-doped anatase and rutile as calculated by the GGA+U functional as implemented in VASP presented in the range of the experimental bandgap of 3.2 eV. The origin of the Fermi level axis is assumed to be at the valence band maximum. The dotted vertical pink lines indicate the position of the theoretical conduction band minimum. Dotted green lines reperesent unstable charge states.

### C. The Origin of Discrepancies between GGA and GGA+U derived results

A comparison between Fig. 2 and Fig. 3, reveals a discrepancy in the predicted behaviour of Cu dopants probed by the two methods. By using GGA functional, we predicted the stabilization of substitutional Cu dopants in a triply negatively charged state for $E_{Fermi}$ values near CBM. The physical interpretation of $Cu_{Ti}^{3-}$, is a triple acceptor in a semiconductor. Here the charge neutrality requirement is fulfilled by the creation of 3 valence band holes. Such a situation occurs when a $Cu^+$ ion replaces a $Ti^{4+}$ and three holes are attracted to the Cu for charge compensation. This point can be verified by examining the total and partial density of states (P/DOS) of substitutional Cu in both a- and rutile TiO$_2$ as presented in Fig. 4. In both cases, empty Cu 3d states are located close to the VBM and the Fermi level crosses a region with non-zero DOS. Filled Cu 3d states are spread throughout the valence band and hybridize extensively with O 2p states. Such hybridization is characteristic of covalent bonding between Cu and its surrounding O ligands. The covalency of the Ti and O bonds can be confirmed by Mullikan population analysis. O ions were found to bear a net charge of ~ -0.87 e while the Ti ions had a charge of ~ 1.68 e. This is less than the half of the nominal charge in fully ionic TiO$_2$ and suggests covalency. A consequence of the hybridization between Cu 3d and O 2p state as predicted by GGA formalism is the lack of magnetic moment for Cu ions, thus TiO$_2$:Cu would not be a magnetic system.

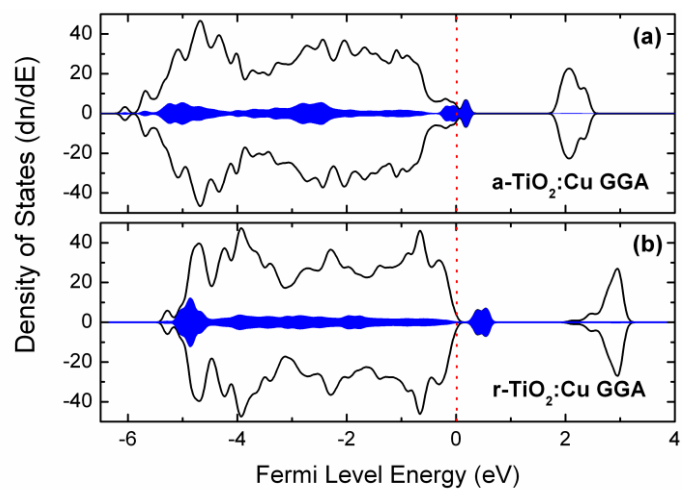

Fig. 4. Total and partial density of states of (a) anatase TiO$_2$ and (b) rutile TiO$_2$ doped with substitutional Cu. Solid black lines represent total density of states while the blue shaded area represents Cu 3d states. The valence band is mainly formed from O 2p states while the conduction band in mainly formed from Ti 3d states.

Through the use of a Hubbard term, the GGA+U formalism produces substantially different band structure for the TiO$_2$:Cu system. As shown in Fig.s 5(a) and (b) in both a- and r-TiO$_2$, Cu's filled 3d states are deeply localized at the bottom of the valence band, thus 3d-2p hybridization is greatly reduced relative to findings obtained using GGA alone. This description is a characteristic of a highly ionic bonding between Cu ion and O ligands. Cu's empty states are now located in the band gap region near the Fermi level while the conduction band is located at a higher level thus increasing the fundamental band gap to ~3 eV. A Mullikan population analysis indicates that the local charge of O ions is -1.98 e while that of Ti ions is 4.22 e. One consequence of highly localized electronic states, predicted by GGA+U formalism, is the local magnetic moment of Cu ions that was absent from the description provided by GGA formalism. Here Mullikan population analysis predicts a local net spin population at Cu sites of -0.49 e and 0.39 e in the anatase and rutile $TiO_2$ respectively.

When compared to experimental findings, the GGA+U formalism gives a closer description of observed behaviour. The reported low values of conductivity in TM doped $TiO_2$ systems are consistent with an ionic system[31]. Furthermore, the difference between the Pauling electronegativity of Ti and O ions is 1.9[32] and this value is indicative of ionic bonding between Ti and O, consistent with the results from the GGA+U formalism. Thus the physical interpretation of $Cu_{Ti}^+$ in the $TiO_2$ host lattice is a fully ionized $Cu^+$ copper ion located in a matrix of $Ti^{4+}$ and $O^{2-}$ ions. In ionic solids charge neutrality of substitutional dopants is usually maintained by the formation of Schottky pairs, for this reason we expect lower concentrations of the mobile holes than that predicted by the GGA method.

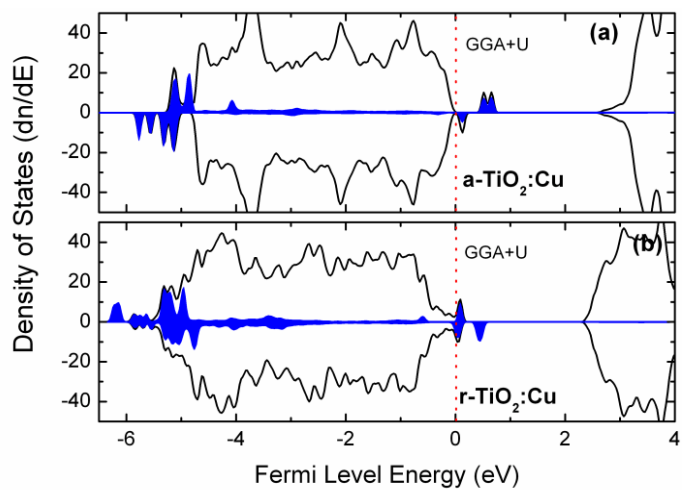

Fig. 5. Total and partial density of states of (a) anatase $TiO_2$ and (b) rutile $TiO_2$ doped with substitutional Cu obtained by GGA+U formalism. Solid black lines represent total density of states while the blue shaded area represents Cu 3d states. Those Cu 3d peaks within the band gap denote Cu 3d empty states that form an impurity band near the VBM edge.

### IV. MAGNETIC INTERACTIONS IN CU DOPED TIO₂

We have established that GGA+U provides a more accurate account of the electronic and bonding behaviour in the TiO$_2$:Cu system. Therefore, the magnetic behaviour of the Cu-TiO$_2$ system was studied only with GGA+U formalism. Since in ionic TiO$_2$ Cu takes a charge state of 1+, the presence of two Cu ions limits the net charge of the supercell to 2. For this reason the magnetic interactions among Cu ions was studied for charges between 0 and 2. Table 1 shows ΔE values for anatase and rutile systems of all considered charge states doped in the compact and scattered configurations illustrated in Fig. 1 (ΔE=E$_{AFM}$- E$_{FM}$). Positive ΔE values indicate the stability of parallel spin alignment among 3d electrons of Cu ion and consequent dominant ferromagnetic behaviour of the TiO$_2$:Cu system. Using ΔE values, the upper limit of Curie temperatures can be calculated using the Heisnberg model with mean field approximation given by the following expression:[33]

$$T_C^{MF} = 12 J_{Cu-Cu} \frac{S_{Cu}(S_{Cu}+1)}{3k_b} \quad (1)$$

Where $J_{Cu-Cu}$ is the effective exchange parameter between Cu dopants, $S_{Cu}$ is the total spin of Cu$^+$ ion and $k_B$ is the Boltzmann constant. The values of $S_{Cu}$ are extracted from the d shell Mullikan electronic population of the Cu ions and $J_{Cu-Cu}$ is assumed to be $J = -\Delta E/12$.[34] ΔE, E$_b$ and T$_c$ of all studied systems is presented in Table 1.

In anatase TiO$_2$:Cu systems the ΔE values for compact values increase from 51 meV for charge state of zero, to 76 meV for a charge state of 1+ and 71 meV for a charge state of 2+. This trend implies that magnetism in enhanced by increasing the charge of the system or in other words, by electron doping. Furthermore the highest value of ΔE in a-TiO$_2$ is 179 meV, found for a scattered dopant configuration with a charge state of 2+ , implying long range magnetic interactions. This is characteristic of RKKY magnetism where the magnitude of the magnetic exchange oscillates periodically with TM-ion-separation.[35] Such a high value for ΔE translates to a T$_c$ of 1124 K.

In contrast to anatase TiO$_2$, rutile TiO$_2$:Cu systems are found to exhibit ferromagnetic behaviour only in compact dopant configurations with ΔE values of 145 meV, 98 meV and 130 meV for charge states of 0, 1+ and 2+ respectively, suggesting a short-range magnetic interaction mediated by nearest neighbour superexchange.[36] This suggests long range magnetism in Cu doped r-TiO2 is not attainable. This is in contrast to the V doped r-TiO2 in which the magnetic interaction among V ions persisted beyond nearest neighbour configuration.[22]

Finally, negative binding energy ($E_b$) values for all charge states of a-TiO$_2$ indicate that compact dopant configurations are more stable in anatase polymorph, illustrating the tendency of copper dopants to segregate. This tendency implies the formation of dopant complexes and a secondary CuO phase at sufficiently high Cu concentrations and temperatures. Noticeably, for r-TiO2 supercell with charge of 2+, where both Cu ions are in the +1 state, $E_b$ is positive for the rutile phase implying no complex formation in contrast to the anatase phase. The tendency for dopant segregation indicates that appropriate fabrication methods are required to avoid the formation of dispersed non-magnetic copper oxide phases in anatase.

## V. SUMMARY

It was found that DFT calculations using GGA+U formalism give a meaningful description of the electronic structure of TiO$_2$:Cu systems. Cu dopants were predicted to exhibit dilute bound-polaron-based ferromagnetic behaviour above room temperature. The TiO$_2$:Cu system is concluded to be a good candidate diluted magnetic semiconductor material. While rutile TiO$_2$:Cu is predicted to function as a DMO only for compact dopant configurations, copper dopants in the metastable anatase phase exhibit stable long range dilute-ferromagnetism and hence this TiO$_2$ polymorph is the better candidate for DMO devices. The tendency for dopants to aggregate in anatase and the absence of long range ferromagnetism in rutile suggests that at elevated temperatures a decay in the diluted magnetic behaviour of a-TiO$_2$:Cu systems may occur as the result of the formation of a secondary CuO phase or the transformation of metastable anatase to the equilibrium rutile phase.

## ACKNOWLEDGMENT

Computational facilities were provided by INTERSECT NSW through project db1.

Table 1. $\Delta E$, $E_b$ and $T_c$ for the a- and r-TiO$_2$ systems in both compact and scattered configurations are given. $\Delta E$ and $E_b$ indicate the nature of magnetic interactions and the tendency toward aggregation respectively while $T_c$ is the theoretical upper limit for Curie temperature.

| System | Supercell Charge (e) | $\Delta E$ (meV) | $E_b$ (meV) | $T_c$ (K) |
|---|---|---|---|---|
| Compact a-TiO$_2$ | 0 | 51 | -113 | 395.2 |
|  | 1 | 76 | -148 | 589.8 |
|  | 2 | 71 | -146 | 548.8 |
| Scattered a-TiO$_2$ | 0 | 42 | — | 1124.3 |
|  | 1 | 64 | — | 759.6 |
|  | 2 | 179 | — | 1003.5 |
| Compact r-TiO$_2$ | 0 | 145 | -91 | 323.1 |
|  | 1 | 98 | -167 | 496.4 |
|  | 2 | 130 | 48 | 1381.0 |
| Scattered r-TiO$_2$ | 0 | -575 | — | — |
|  | 1 | -47 | — | — |
|  | 2 | -48 | — | — |

1. J. M. D. Coey, Curr. Opin. ST. M. **10,** 83 (2006).
2. S. J. Pearton, C. R. Abernathy, D. P. Norton, A. F. Hebard, Y. D. Park, L. A. Boatner and J. D. Budai, Mat. Sci. Eng. R **40,** 137 (2003).
3. S. A. Wolf, D. D. Awschalom, R. A. Buhrman, J. M. Daughton, S. von Molnár, M. L. Roukes, A. Y. Chtchelkanova and D. M. Treger, Science **294,** 1488 (2001).
4. T. Dietl, H. Ohno and F. Matsukura, Phys. Rev. B **63,** 195205 (2001).
5. J. M. D. Coey, M. Venkatesan and C. B. Fitzgerald, Nat. Mater. **4,** 173 (2005).
6. Y. Matsumoto, M. Murakami, T. Shono, T. Hasegawa, T. Fukumura, M. Kawasaki, P. Ahmet, T. Chikyow, S.-y. Koshihara and H. Koinuma, Science **291,** 854 (2001).
7. D. A. H. Hanaor and C. C. Sorrell, J. Mater. Sci. **46,** 855 (2011).
8. S. B. Ogale, Adv. Mater. **22,** 3125 (2010).
9. T. Dietl, Nat. Mater. **9,** 965 (2010).
10. G. Talut, H. Reuther, J. Grenzer and S. Zhou, *Origin of ferromagnetism in iron implanted rutile single crystals*. (Springer, New York, 2009).
11. K. A. Griffin, A. B. Pakhomov, C. M. Wang, S. M. Heald and K. M. Krishnan, Phys. Rev. Lett. **94,** 157204 (2005).
12. S. A. Chambers, T. Droubay, C. M. Wang, A. S. Lea, R. F. C. Farrow, L. Folks, V. Deline and S. Anders, Appl. Phys. Lett. **82,** 1257 (2003).
13. S. A. Chambers, Surf. Sci. Rep. **61,** 345 (2006).
14. K. Sato, L. Bergqvist, J. Kudrnovský, P. H. Dederichs, O. Eriksson, I. Turek, B. Sanyal, G. Bouzerar, H. Katayama-Yoshida, V. A. Dinh, T. Fukushima, H. Kizaki and R. Zeller, Rev. Mod. Phys. **82,** 1633 (2010).
15. G. Cohen, V. Fleurov and K. Kikoin, J. Appl. Phys. **101,** 09H106 (2007).
16. C. J. Cramer and D. G. Truhlar, Phys. Chem. Chem. Phys. **11,** 10757 (2009).
17. S.-D. Mo and W. Y. Ching, Phys. Rev. B **51,** 13023 (1995).
18. T. Umebayashi, T. Yamaki, H. Itoh and K. Asai, J. Phys. Chem. Solids **63,** 1909 (2002).
19. R. Asahi, T. Morikawa, T. Ohwaki, K. Aoki and Y. Taga, Science **293,** (2001).
20. G. Pacchioni, J. Chem. Phys. **128,** 182505 (2008).
21. A. Walsh, J. L. F. Da Silva and S.-H. Wei, Phys. Rev. Lett. **100,** 256401 (2008).
22. J. Osorio-Guillen, S. Lany and A. Zunger, Phys. Revi. Lett. **100,** 036601 (2008).
23. B. J. Morgan and G. W. Watson, Surf. Sci. **601,** 5034 (2007).
24. B. Delley, J. Chem. Phys. **92,** 508 (1990).
25. B. Delley, J. Chem. Phys. **113,** 7756 (2000).
26. D. A. H. Hanaor, M. H. N. Assadi, S. Li, A. Yu and C. C. Sorrell, Comput.l Mech. **50,** 185 (2012).
27. G. Kresse and J. Furthmuller, Phys. Rev. B **54,** 11169 (1996).
28. J. P. Perdew, K. Burke and M. Ernzerhof, Phys. Rev. Lett. **77,** 3865 (1996).
29. R. Neudert, M. Knupfer, M. S. Golden, J. Fink, W. Stephan, K. Penc, N. Motoyama, H. Eisaki and S. Uchida, Phys. Rev. Lett. **81,** 657 (1998).
30. N. H. Vu, H. V. Le, T. M. Cao, V. V. Pham, H. M. Le and N.-M. Duc, J. Phys.: Condens. Mat. **24,** 405501 (2012).
31. R. Janisch, P. Gopal and N. A. Spaldin, J. Phys.: Condens. Mat. **17,** R657 (2005).
32. M. D. Segall, R. Shah, C. J. Pickard and M. C. Payne, Phys. Rev. B **54,** 16317 (1996).
33. G. Rushbrooke and P. J. Wood, Mol. Phys. **1,** 257 (1958).
34. T. Chanier, M. Sargolzaei, I. Opahle, R. Hayn and K. Koepernik, Phys. Rev. B **73,** 134418 (2006).
35. D. J. Priour, Jr., E. H. Hwang and S. Das Sarma, Phys. Rev. Lett. **92,** 117201 (2004).
36. P. W. Anderson, Phys. Rev. **115,** 2 (1959).